\documentclass[preprint,5p,twocolumn,review]{elsarticle}

\usepackage{amssymb}
\usepackage{amsmath}
\usepackage{graphicx}
\graphicspath{{pics/}}
\usepackage{booktabs}
\usepackage{multirow}
\usepackage{subfigure}
\usepackage{amssymb}
\usepackage{xcolor}
\usepackage[switch]{lineno}

\let\oldequation\equation
\let\oldendequation\endequation

\renewenvironment{equation}
  {\linenomathNonumbers\oldequation}
  {\oldendequation\endlinenomath}

\begin{document}

\begin{frontmatter}



\title{Circuit Solutions towards Broadband Piezoelectric Energy Harvesting: An Impedance Analysis}


\author[inst1,inst2]{Bao Zhao\corref{mycorrespondingauthor}}
\author[inst1]{Junrui Liang\corref{mycorrespondingauthor}}
\cortext[mycorrespondingauthor]{Corresponding author: bao.zhao@ibk.baug.ethz.ch (Bao Zhao); liangjr@shanghaitech.edu.cn (Junrui Liang)}

\nonumnote{This article was presented in part at the 2018 SPIE Smart Structures + Nondestructive Evaluation Conference \cite{Zhao2018a}. }
\affiliation[inst1]{organization={School of Information Science and Technology, ShanghaiTech University},
            city={Shanghai},
            country={China}}

\affiliation[inst2]{organization={Department of Civil, Environmental, and Geomatic Engineering, ETH Zurich},
            city={Zurich},
            country={Switzerland}}

\begin{abstract}
In the studies of piezoelectric energy harvesting (PEH) systems, literature has shown that circuit advancement has a significant effect on the enhancement of energy harvesting capability in resonance. On the other hand, some recent studies using the phase-variable (PV) synchronized switch technologies have found that the advanced circuit solutions can also broaden the harvesting bandwidth, i.e., improving the off-resonance energy harvesting capability, to some extent. However, the available span of the electrically induced dynamics by the existing energy harvesting circuits was not properly defined and demonstrated. Performance comparison among different circuits cannot be fairly achieved without using a common theoretical language. Given these, this paper provides an impedance-based analysis and comparison on the electromechanical joint dynamics of the PEH systems using different interface circuits, in particular, with a specific focus on their contributions towards the improvement of energy harvesting bandwidth. Given that the resonance tunability by circuit solutions has received no attention in the conventional ideal model of kinetic energy harvester, we firstly propose a more inclusive ideal model for better generalization. In practice, it was proven that the attainable dynamic ranges of the practical energy harvesting circuits are only some subsets of the ideal realm. A detailed quantitative study on the attainable ranges of the PV synchronized switch circuit solutions is provided after the introduction of the ideal target. Simulation and experimental results of different interface circuits show good agreement with the theoretical analysis. It can be concluded that the resonance tunability strongly depends on the achievable extent in the reactive (imaginary) direction of the equivalent impedance plane. In practice, the electromechanical coupling conditions and dielectric loss might also influence the resonance tunability. The general ideal model and quantitative impedance analysis provided in this paper help guide the future design effort towards high-capability and broadband PEH systems.
\end{abstract}





\end{frontmatter}

	\section{Introduction}
	\label{sec:intro}

	In recent years, the kinetic energy harvesting (KEH) systems have caught more research interests, with the purpose of enabling some highly distributed Internet of Things (IoT) devices, which operate in vibrational environments, to become energy self-sufficient. No matter what mechanical design, transducer, or power conditioning circuit are used, there are two most significant design targets for the KEH systems.
	\begin{enumerate}
		\item \textit{The high-capability (HC) target}: to increase the harvested power in resonance.
		\item \textit{The broadband (BB) target}: to increase the off-resonance harvested power, i.e., to broaden the harvesting bandwidth.
	\end{enumerate}
	For the KEH systems using piezoelectric transducers, i.e., piezoelectric energy harvesting (PEH) systems, many mechanical or electrical solutions were proposed regarding to the aforementioned two design targets.

	For the HC design target, the mechanical solutions include: to increase the amount of active materials, e.g., using a bimorph instead of a unimorph for energy harvesting \cite{Ng2005}; or to decrease the equivalent mechanical stiffness, e.g., applying an axial load to the piezoelectric beam \cite{Cottone2012,wang2019integration,cha2019parameter}. All mechanical solutions have led to the increase of the system electromechanical coupling coefficient. On the other hand, without changing the mechanical structure, it was shown that the harvesting capability could be effectively enhanced by advancing the interface circuit designs \cite{Liang2017, Szarka2012}. The featured interface circuits include the synchronous electric charge extraction (SECE) \cite{Lefeuvre_Badel_Richard_Guyomar_2005}, synchronized switch harvesting on inductor (SSHI) \cite{Guyomar2005}, single-supply pre-biasing (SSPB) \cite{Dicken_Mitcheson_Stoianov_Yeatman_2012}, and synchronized triple bias-flip (S3BF) \cite{zhao2020series}. Shu et al. have intuitively put that, by using the synchronized switch interface circuits, a weakly coupled PEH system might become a moderately or strongly coupled system \cite{Shu2007}.

	For the BB design target, most research efforts came from the mechanical community. The major mechanical solutions include: combining multiple vibrators at different resonant frequencies \cite{zhao2022ecm}; tuning the vibration modes by adding auxiliary structures \cite{zhou2019new}; and introducing the nonlinear mechanical dynamics to the linear vibrator \cite{Tang2010, yuan2019harmonic, fang2022multistability}. Considering a PEH system as an organic electromechanical system, each part of the system should have some effect on the global dynamic characteristics. In this sense, the circuit solutions should also play a role in the system dynamics. However, given the major attention on the HC target for weakly coupled systems, few circuit studies realized the BB target until recent years \cite{morel2022comparative}. The dynamic contribution of an interface circuit comes from the inverse piezoelectric effect. Therefore, the dynamic intervention can be more obviously observed under strong coupling \cite{Morel2019,gibus2022non}. Without a strong coupling system, the connected interface circuit has little to do with the mechanical vibration. By modifying the synchronized switch instants in SSHI, which tunes the weakly coupled systems into strongly coupled ones, Hsieh et al. have proposed the phase-variable (PV) SSHI towards the BB target \cite{Hsieh2015}. A similar idea was also presented in a US Patent filed by Li et al. \cite{Li2016}. Lefeuvre et al. have introduced the PV control to the SECE interface circuit for the same BB purpose \cite{Lefeuvre2017, Lefeuvre2017a}. Zhao et al. also utilized the PV control in the P-S3BF interface circuit for the BB target in a nonlinear energy harvesting system \cite{zhao2020dual}. These technologies are referred to as PV-SSHI, PV-SECE , and PV-P-S3BF in the following part of this paper. Besides the phase-variable control, Brenes et al. \cite{brenes2020large} and  Morel et al. \cite{morel2018frequency} also include additional tunable parameters such as synchronized switch duration from frequency tuning (FT) SECE \cite{brenes2020large} and short circuit phase from short circuit (SC) SECE \cite{morel2018frequency} to further increase the PEH bandwidth.

	In general, as two constitutive parts of a PEH system, both the mechanical or electrical designs act some roles towards the HC and BB targets. On the other hand, the exploration of circuit solutions towards the BB target has just began.

	For formulating the dynamics of various designs, mechanical experts tend to use a \textit{top-down solution}. They start from the comprehensive equations of motion and finally arrive at the closed-form expressions of displacement, harvested power, etc. \cite{Guyomar2005, Shu2007}. On the contrary, electrical engineers are more used to the modular way of thinking and tend to use a \textit{bottom-up solution}. They derive the dynamic expressions of all modules in a system, substitute the corresponding symbols in the system-level equations, and finally obtain the numerical results of displacement, harvested power, etc. \cite{Kong2010, Liang2012}. For the studies of dynamic PEH systems using different interface circuits, it has been shown that both the top-down and bottom-up solutions lead to very close results \cite{Shu2007, Liang2012}. Moreover, by taking the modular way of thinking, the bottom-up solution has two additional merits:
	\begin{itemize}
		\item The dynamics of different interface circuits can be singled out and intuitively illustrated and compared.
		\item The system-level equations can be reused regardless of the specific connected interface circuit.
	\end{itemize}
	The equivalent impedance modeling has demonstrated the aforementioned merits in PEH system analysis \cite{Liang2012, Liang2014, Chen2019}. This paper extends the benefit of impedance analysis for generating a more comprehensive understanding of the conventional HC oriented SSHI, SECE, and the cutting-edge BB oriented PV-SSHI and PV-SECE technologies, including their similarities and differences, effective conditions, limitations, etc.

	The paper is organized as follows. Section I explains the HC and BB design targets for PEH systems and the necessity of impedance analysis for the cutting-edge interface circuit solutions. Section II introduces the general ideal model towards HC and BB targets. Section III briefly reviews the working principle of the PV-SECE and PV-SSHI solutions. Section IV derives their equivalent impedance model. Section V experimentally validates the theoretical analysis. Section VI concludes the paper.

	\section{Models}
	\label{sec:model}

	Both the inventions of the PV-SECE and PV-SSHI technologies have brought in the idea of resonance tuning in an electrical way for PEH systems. Nevertheless, such idea was not thoroughly elaborated in a general context. It is necessary to clarify the design targets by referring to the general ideal model of a KEH system.

	\subsection{Ideal model for resonance-untunable systems}

	\begin{figure}[!t]
		\centering
		\includegraphics[width=1\columnwidth]{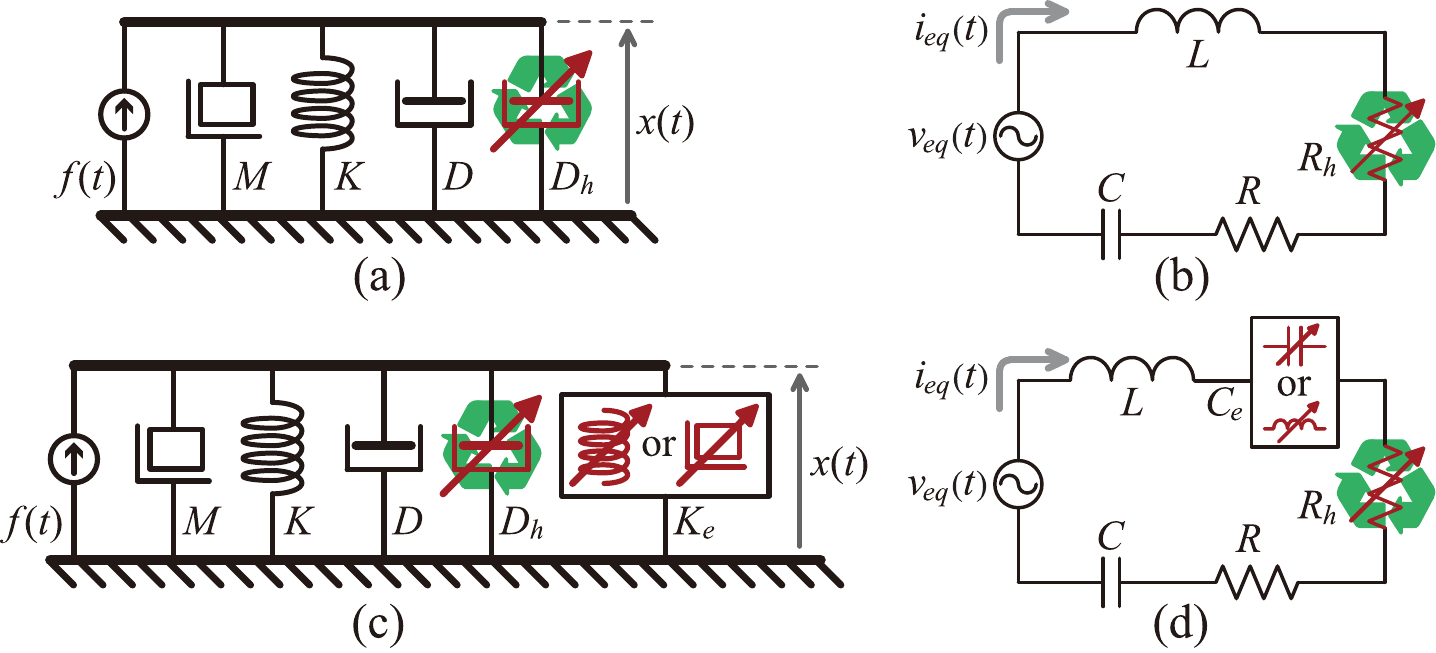}
		\caption{The ideal model of a KEH system. (a) The conventional ideal model only considering $D_h$ the energy harvesting induced damping \cite{WilliamsandYates1996}. (b) The corresponding equivalent circuit of the conventional ideal model in the PEH cases. (c) The general ideal model including $K_e$ the electrically induced stiffness (positive number) or mass (negative number). (d) The corresponding equivalent circuit of the general ideal model in the PEH cases.}
		\label{fig:PEH_model}
	\end{figure}

	The most referred ideal model for KEH systems was introduced by Williams and Yates in 1996 \cite{WilliamsandYates1996, Mitcheson2008, Beeby2006, Cook-Chennault2008}. Fig. \ref{fig:PEH_model}(a) shows the mechanical schematics of such ideal model. The idea KEH system is composed of a linear lumped mechanical vibrator, which is characterized by its mass $M$, stiffness $K$, and damping coefficient $D$, and an additional tunable energy harvesting induced damping coefficient\footnote{$D_h$ was referred to as the electrically induced damping in \cite{WilliamsandYates1996}. For better differentiating the dissipative damping and regenerative damping, we specify $D_h$ as the energy harvesting induced damping and decorate its icon with a recycling symbol in Fig. \ref{fig:PEH_model}.} $D_h$. The applied force and displacement of energy harvester are denoted as $f(t)$ and $x(t)$ in Fig. \ref{fig:PEH_model}(a). For an inertial vibrator under base excitation, we have $f(t)=-M\ddot{y}(t)$, where $\ddot{y} (t)$ is the base acceleration and $x(t)$ becomes the relative displacement of the vibrator \cite{WilliamsandYates1996}.

	Besides the mechanical representation, the ideal KEH system can also be presented in an electrical way, as shown in Fig. \ref{fig:PEH_model}(b). For example, in a PEH system, by taking the electromechanical analogy, the dynamics of $M$, $K$, $D$ in the mechanical domain can be equivalently expressed with inductance $L$, capacitance $C$, and resistance $R$ with the relations as follows:
	\begin{equation}
	R={D}/{\alpha^{2}},\qquad
	C={\alpha^{2}}/{K},\qquad
	L={M}/{\alpha^{2}},
	\label{eq:RLC}
	\end{equation}
	where $\alpha$ is the force-voltage factor in the piezoelectric coupled system. In the piezoelectric systems, The effort and flow variables in the electrical domain correspond to the flow and effort variables in the mechanical domain with the relations as follows:
	\begin{equation}
	{{v}_{eq}}(t)={f(t)}/{{{\alpha }}}, \qquad
	{{i}_{eq}}(t)={\alpha }\dot{x}(t),
	\end{equation}
	where $v_{eq}(t)$ and $i_{eq}(t)$ are the equivalent voltage and current, respectively; $\dot{x}(t)$ is the vibration velocity. In the electrical equivalent network, the regenerative damping effect is represented by a tunable regenerative resistance $R_h$, as shown in Fig. \ref{fig:PEH_model}(b). The equivalent impedance network of the KEH systems using other transduction mechanisms might be different, say, it is a series RLC network driven by a current source in the electromagnetic cases \cite{Liang2017a}. In this paper, we focus on the PEH cases.

	Given a KEH system under harmonic base excitation, the governing equation of the ideal KEH system is formulated as follows:
	\begin{equation}
	M\ddot x\left( t \right) + \left( {{D} + {D_h}} \right)\dot x\left( t \right) + Kx\left( t \right) =  - M\ddot y\left( t \right).
	\label{eq:1}
	\end{equation}
	In the frequency domain, \eqref{eq:1} can be solved as follows:
	\begin{equation}
	X\left( {\omega } \right) = \frac{{{{\tilde \omega }^2}}}{{1 - {{\tilde \omega }^2} + j2 \tilde \omega \left( {1 + \eta } \right){\zeta }}}Y\left( {\omega } \right).
	\end{equation}
	$\tilde\omega=\omega/\omega_n$ is the normalized frequency, where $\omega$ is the operating frequency; $\omega_n=\sqrt{K/M}$ is the natural frequency. $\eta=\zeta_h/\zeta=D_h/D$ denotes the ratio between energy harvesting induced damping and inherent mechanical damping, where ${\zeta} = {D}/(2M\omega_n)$ and ${\zeta_h}=D_h/(2M\omega_n)$ are their corresponding damping ratios.

\begin{figure}[!t]
		\centering
		\includegraphics[width=1\columnwidth,page=2]{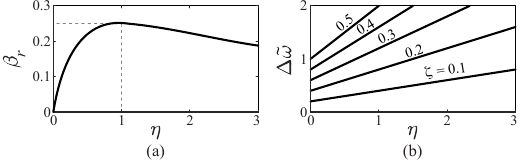}
		\caption{(a) The normalized harvested power in resonance. (b) The normalized half-power bandwidth.}
		\label{fig:oldbandwidth}
	\end{figure}

	The maximum power absorbed by the mechanical damping component $D$ is obtained when there is no energy harvesting induced damping and the system vibrates in resonance, i.e.,
	\begin{equation}
	P_{m,\text{max}} =\left.\frac{D\omega^2 |X(\omega)|^2}{2}\right|_{\eta = 0,\,\tilde\omega=1} = \frac{M^2A_Y^2}{{2D}},
	\label{eq:Pm_max}
	\end{equation}
	where ${A_Y}={\omega_n^2}\left| Y(j\omega) \right|$ is the base acceleration magnitude. On the other hand, the harvested power absorbed by $D_h$ can be formulated as follows:
	\begin{equation}
	P_h = \beta_o \beta _r P_{m,\text{max}},
	\end{equation}
	where
	\begin{equation}
	{\beta _r} =\frac{\left.P_h\right|_{\tilde\omega=1}}{P_{m,\text{max}}}= \frac{\eta }{{{{\left( {1 + \eta } \right)}^2}}},
	\end{equation}
	is the power ratio between maximum harvested power in resonance and the maximum mechanical damping power in \eqref{eq:Pm_max};
	\begin{equation}
	{\beta _o} =\frac{P_h}{\left.P_h\right|_{\tilde\omega=1}}= \frac{[2\tilde{\omega}(1+\eta)\zeta]^2}{(1-\tilde{\omega}^2)^2+[2\tilde{\omega}(1+\eta)\zeta]^2}
	\label{eq:beta_o}
	\end{equation}
	is the normalized power absorbed by the energy harvesting introduced damping component $D_h$ under the normalized frequency $\tilde{\omega}$. $\beta_r$ attains its maximum when $\eta=1$, i.e., $D_h=D$. In this case, we can have the maximum $\beta_r=1/4$, as shown in Fig. \ref{fig:oldbandwidth}(a). In resonance, the maximum harvested power can be expressed as follows:
	\begin{equation}
	P_{h,\text{max}}=\frac{1}{4}P_{m,\text{max}}=\frac{M^2A_Y^2}{{8D}}=\frac{F^2}{{8D}},
	\label{eq:Pe_max}
	\end{equation}
	where $F$ denotes the force magnitude. Equations \eqref{eq:Pm_max} and \eqref{eq:Pe_max} repeat the conclusion, which was drawn in \cite{WilliamsandYates1996}. Maximum harvested power is obtained by matching the mechanical inherent damping $D$ with a tunable energy harvesting induced damping $D_h$. However, in the previous studies, the energy harvested bandwidth of the conventional ideal model was not studied. Such bandwidth can be evaluated based on the expression of normalized power $\beta_o$ in \eqref{eq:beta_o} as the distance between two half-power points, i.e., the roots of the equation $\beta_o=1/2$. Denoting the two positive roots as $\tilde \omega_{1}$ and $\tilde \omega_{2}$, the normalized half-power bandwidth can be formulated as follows
	\begin{equation}
	\Delta \tilde \omega  = \frac{{\Delta \omega }}{{{\omega _n}}} = \left| {{{\tilde \omega }_{1}} - {{\tilde \omega }_{2}}} \right| = 2\left( {1 + \eta } \right){\zeta}.
	\end{equation}
	Fig. \ref{fig:oldbandwidth}(b) illustrates the changing trend of $\Delta \tilde \omega$ under different $\eta$ and $\zeta$. It can be summarized that, no matter how much is $\zeta$ the mechanical damping ratio, the half-power bandwidth increases as $\eta$ the ratio between energy harvesting induced damping and mechanical inherent damping gets larger.

	The two sub-figures in Fig. \ref{fig:oldbandwidth} give an idea about the HC and BB performance of the conventional ideal model, i.e., the ideal KEH system without resonance tunability. For the $\eta<1$ cases, which are the characteristics of most PEH implementations, in particular, weakly coupled systems, increasing the energy harvesting induced damping not only leads to higher energy harvesting capability \cite{Guyomar2005}, as shown in Fig. \ref{fig:oldbandwidth}(a), but also slightly broadens the energy harvesting bandwidth \cite{Shu2007}, as shown in Fig. \ref{fig:oldbandwidth}(b).

	\subsection{Ideal model for resonance-tunable systems}

\begin{figure}[!t]
		\centering
		\includegraphics[width=0.8\columnwidth,page=3]{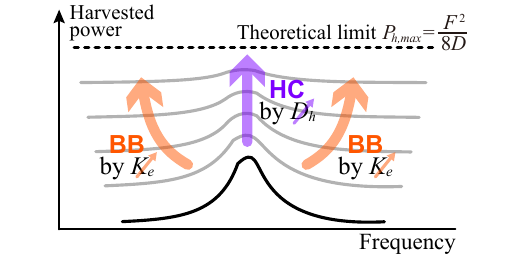}
		\caption{The harvested power versus frequency relation and the visualization of HC and BB design targets.}
		\label{fig:HC_BB}
	\end{figure}

	The BB targets can be better realized if the harvesting circuit, as a constitutive part of the electromechanical system, can have some resonance tunability. Given this, Fig. \ref{fig:PEH_model}(c) shows a more general KEH system by adding a tunable reactive component $K_e$ to the conventional model. $K_e$ is tunable as $D_h$ did. It can be either a tunable stiffness or a tunable mass. The corresponding equivalent circuit is shown in Fig. \ref{fig:PEH_model}(d). Assuming the unlimited and arbitrary tunability for $D_h$ and $K_e$, $P_{h,\text{max}}$ can be always attained by making $D_h=D$ and $K_e+K=\omega^2M$, i.e., the \textit{conjugate impedance matching} condition, given the source and load impedances as
	\begin{equation}
	\left\{
	\begin{aligned}
	&Z_{\text{source}}(\omega)=D+j(\omega M-K/\omega),\\
	&Z_{\text{load}}(\omega)=D_h-jK_e/\omega,
	\end{aligned}
	\right.
	\label{eq:matching}
	\end{equation}
	respectively. Putting aside the slight bandwidth broadening effect realized by the real-part matching condition, it can be roughly taken that the $D_h$ and $K_e$ adjustments are related to the HC tuning in resonance and BB tuning off resonance, respectively. Fig. \ref{fig:HC_BB} illustrates the tuning effort towards $P_{h,\text{max}}$ the theoretical limit of harvested power, which can be broken down into the HC and BB targets.

	\subsection{General model for practical systems}

	\begin{figure}[!t]
		\centering
		\includegraphics[width=1\columnwidth,page=4]{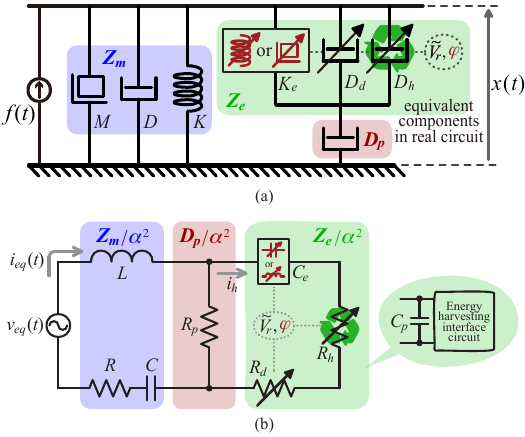}
		\caption{The general model of a practical KEH system (two-parameter tunable). (a) The mechanical schematics. (b) The electrical equivalent impedance network in the PEH cases.}
		\label{fig:2parameter}
	\end{figure}
	Ideal models help indicate the research direction. However, in practice, the circuit dynamics is not as simple as the aforementioned ideal case \cite{Liang2012, Liang2014}. The nonideality is embodied in all aspects including, but not limited to:
	\begin{itemize}
		\item The electrical induced dynamics is not arbitrarily attainable. It is bounded. Different interface circuits give different constraints.
		\item Dissipative resistance coexists with the regenerative resistance during the power conditioning process. Both of them have the same outward effect on structural damping \cite{Liang2009}.
		\item The HC and BB targets cannot be independently made with the existing energy harvesting interface circuits.
		\item Parasitic characteristics, e.g., the dielectric loss in the piezoelectric transducer, might shrink the attainable dynamic range and therefore weaken the tunability \cite{Liang2014}.
	\end{itemize}
	Given these complexities in practice, it is inappropriate to simply take the well-established impedance matching concept, i.e., making $Z_{\text{load}}=Z_{\text{source}}^*$ in \eqref{eq:matching}, for the KEH power optimization. Such an easy concept only gives a raw idea about the dynamic tunabilities of different interface circuits \cite{Liao_Liang_2018, Liao2019}. A general dynamic model was proposed for the PEH systems, given the aforementioned practical issues \cite{Liang2014}. Fig. \ref{fig:2parameter} shows the general mechanical schematic model and its electrical equivalent. There are some significant differences between the practical equivalent and the ideal models, which are listed as follows
	\begin{itemize}
		\item The dissipative and regenerative damping components are distinguished as $D_d$ and $D_h$ in the mechanical domain (and $R_d$ and $R_h$ in the electrical domain), respectively.
		\item The dependency of the variable components on the tunable circuit parameters, e.g., rectified voltage $\tilde{V}_r$, is explicitly illustrated by the dashed links.
		\item A series damping component $D_p$ is added for characterizing the dielectric leakage of the piezoelectric element.
	\end{itemize}
	The same model was demonstrated to be also effective in the electromagnetic case, where $D_p$ characterizes the effect equivalent series resistance (ESR) of the electromagnetic coil \cite{Liang2017b}. In this paper, a slight complement is made based on the general model of the practical systems. Given the recent progress on the broadband circuit solutions by modifying \textit{synchronized switch phase} $\varphi$ \cite{Hsieh2015, Lefeuvre2017}, $\varphi$ is regarded as an additional tunable parameter towards richer electrically induced dynamics. The dynamic components are divided into three groups: the \textit{mechanical dynamics}, which is highlighted in blue in Fig. \ref{fig:2parameter} and expressed as follows
	\begin{equation}
		Z_m(\omega)=D+j(\omega M-K/\omega);
	\end{equation}
	the \textit{dynamics of the piezoelectric capacitance $C_p$ and interface circuit combination}, which is highlighted in green and expressed as follows
	\begin{equation}
		\begin{aligned}
			&Z_e(\omega,\tilde{V}_r,\varphi)\\=&D_d(\omega,\tilde{V}_r,\varphi)+D_h(\omega,\tilde{V}_r,\varphi)-jK_e(\omega,\tilde{V}_r,\varphi)/\omega;
		\end{aligned}
		\label{eq:Ze_mech}
	\end{equation}
	and the \textit{dielectric loss}, which is highlighted in red and whose effect is summarized by the damping component $D_p$.

	\begin{figure}[!t]
		\centering
		\includegraphics[width=0.8\columnwidth,page=5]{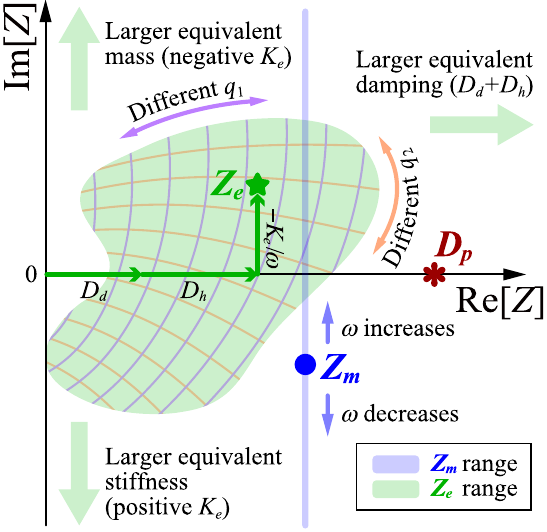}
		\caption{The general impedance picture of the two-parameter ($q_1$ and $q_2$) tunable circuit solutions for PEH systems.}
		\label{fig:generalpic}
	\end{figure}

	\begin{figure*}[!t]
		\centering
		\includegraphics[width=2\columnwidth,page=6]{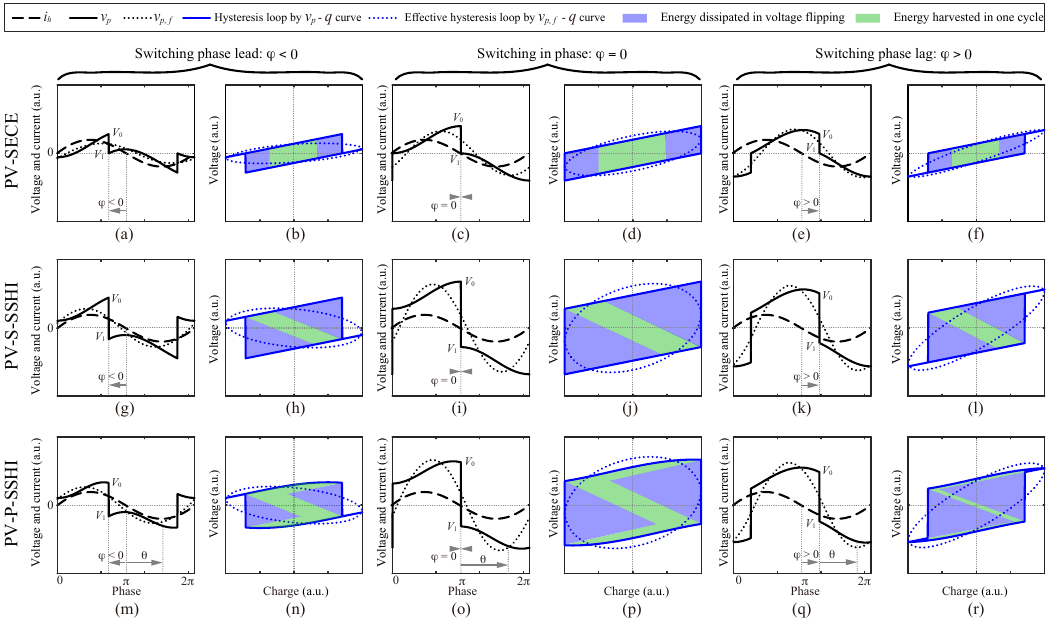}
		\caption{The waveforms and work cycle hysteresis of different circuits under different operations. (a)-(f) PV-SECE. (g)-(l) PV-S-SSHI. (m)-(r) PV-P-SSHI. (a), (b), (g), (h), (m), and (n) $\varphi<0$, the switching phase lead cases. (c), (d), (i), (j), (o), and (p) $\varphi=0$, the in-phase cases, i.e., SECE, S-SSHI, and P-SSHI. (e), (f), (k), (l), (q), and (r) $\varphi>0$, the switching phase lag cases.}
		\label{fig:pvwave}
	\end{figure*}

	\begin{figure}[!t]
		\centering
		\includegraphics[width=1\columnwidth,page=7]{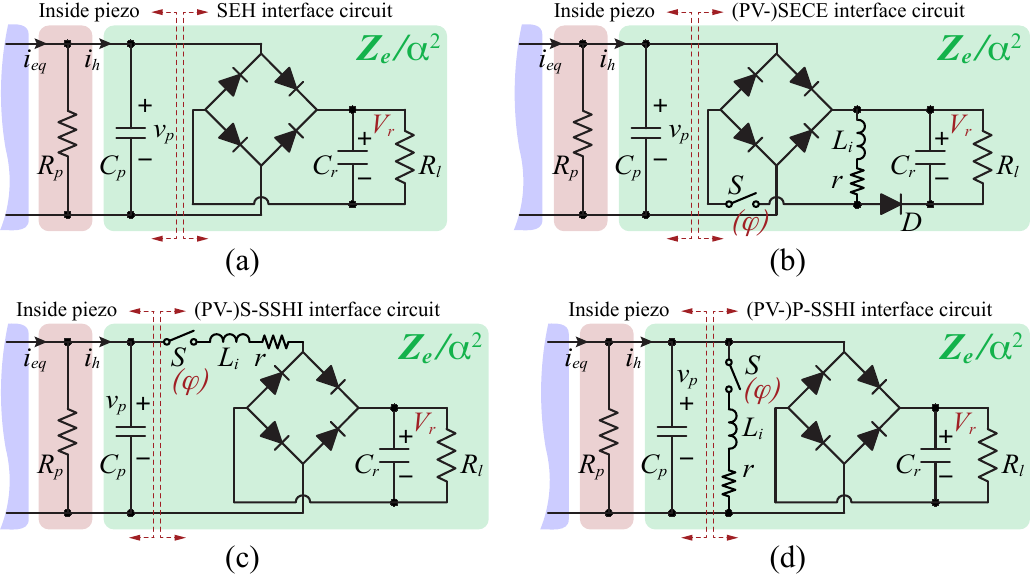}
		\caption{Typical interface circuits for PEH. The mechanical part is the same as that shown in Fig. \ref{fig:2parameter}(b). (a) SEH. (b) (PV-)SECE (it is phase variable when $\varphi$ is tunable). (c) (PV-)S-SSHI. (d) (PV-)P-SSHI.}
		\label{fig:circuits_s}
	\end{figure}

	\begin{figure}[!t]
		\centering
		\includegraphics[width=1\columnwidth,page=8]{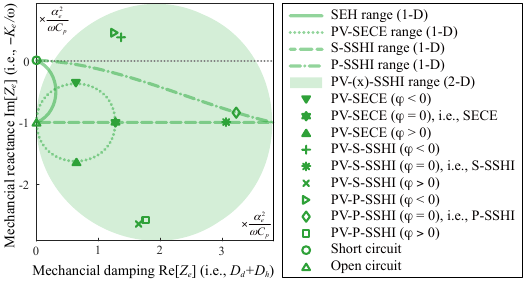}
		\caption{The equivalent impedance picture of different interface circuits under different operations.}
		\label{fig:Ze_s}
	\end{figure}

	Fig. \ref{fig:generalpic} illustrates the dynamic ranges of $Z_m$ (blue), $Z_e$ (green), and $D_p$ (red) on the impedance plane. $Z_m$ moves along the light blue vertical line as the operating frequency $\omega$ changes. It attains the minimum magnitude $D$ (intercept on the real axis) when vibrates in resonance. $D_p$ is a fixed point on the real axis. The most complicated part is $Z_e$. In terms of the functionalities, it can be divided into three dynamic details: the reactive component $-jK_e/\omega$, dissipative damping $D_d$, and regenerative damping $D_h$. The attainable range of $Z_e$ is constrained within some boundaries. If there is no tunable parameter, e.g., SECE, the corresponding $Z_e$ is a fixed point. If there is one tunable parameter, e.g., by taking either the bridge rectifier standard energy harvesting (SEH) interface circuit or SSHI, $Z_e$ can move along a specific 1-D (one-dimensional) curve segment \cite{Liang2012}. If there are two tunable parameters, as generalized by $q_1$ and $q_2$ in Fig. \ref{fig:generalpic}, e.g., by taking either the PV-SSHI \cite{Hsieh2015} or PV-SECE \cite{Lefeuvre2017}, $Z_e$ can move within a specific 2-D (two-dimensional) bounded area. Adding an additional tunable parameter to the interface circuit extends the dynamics tunability.	The development of the one-parameter tunable circuits focuses on the HC target by enlarging the real part of $Z_e$; while the two-parameter tunable circuits can simultaneously change the imaginary part, which enables the advances towards the BB target.

	With the schematic shown in Fig. \ref{fig:2parameter}(a), the harvested power can be formulated as the power extracted by the regenerative damping component $D_h$, i.e.,
	\begin{equation}
	P_h(\omega,\tilde{V}_r,\varphi)=\frac{F^2D_h}{2}\left|\frac{D_p}{Z_m D_p + Z_m Z_e + Z_e D_p}\right|^2,
	\label{eq:Ph}
	\end{equation}
	where $F$ denotes the magnitude of the excitation force. $P_h$ is a function of the operating frequency $\omega$ and two electrically tunable parameters $\tilde V_r$ and $\varphi$. The same result can also be obtained from the electrical equivalent impedance network shown in Fig. \ref{fig:2parameter}(b) \cite{Liang2014}. As we can observe from \eqref{eq:Ph}, given the existence of several functional components and their constraints in practical systems, the problem of the harvested power optimization is more complicated than that described in the conjugate impedance matching case.

	\section{Circuit Solutions towards BB Target}

	The early circuit solutions for PEH enhancement focused on the HC target. After the studies on the baseline SEH solution \cite{Ottman2002}, many interface circuits were proposed for harvesting more power under the same mechanical excitation condition. The most representative interface circuits include SECE, parallel-SSHI (P-SSHI), and series-SSHI (S-SSHI), whose circuit topologies are shown in Fig. \ref{fig:circuits_s}. In the original solutions of SEH, P-SSHI, and S-SSHI, there is only one tunable parameter, usually the rectified voltage $V_r$, which can be mapped from different loading resistance $R_l$. As illustrated in the general picture of Fig. \ref{fig:generalpic}, when there is only one tunable parameter, the equivalent impedance $Z_e$ can merely be adjusted along a specific curve segment. For achieving the HC goal, we have to extend $D_h$ in the real part of $Z_e$, such that it can catch up with the mechanical damping $D$ in weakly or moderately coupled systems. This can be clarified with the impedance analysis of the 1-D tunable circuits \cite{Liang2014, Chen2019}.

	For the BB target, $Z_e$ should reach the region of larger imaginary part. Such goal can be realized by tuning the phase of the switch instants in the latter three circuits in Fig. \ref{fig:circuits_s}(b)--(d). The corresponding solutions were called PV-SECE, PV-P-SSHI, and PV-S-SSHI. The circuit topologies of the PV-x (x stands for the corresponding 1-D tunable solution) circuits remain the same; only the switch control has been changed for realizing the new 2-D tunable solutions.

	\subsection{PV-SECE}

	The PV-SECE solution is based on the original SECE interface circuit proposed by Lefeuvre et al. \cite{Lefeuvre_Badel_Richard_Guyomar_2005}. The SECE circuit can harvest a similar amount of power under different loading conditions. For weakly coupled cases, SECE can harvest three times more power than the basic SEH solution, but less capable than the SSHI solutions under maximum harvesting condition. However, due to the load independence feature of SECE, it doesn't need an additional maximum power point tracking (MPPT) module. Therefore, it has received much attention from the practical point of view.

	The circuit topology of SECE is shown in Fig.  \ref{fig:circuits_s}(b). The interface is composed of a full-wave bridge rectifier and a buck-boost converter. During the SECE operation, the switch $S$ will be turned on for one-fourth of the $L_iC_p$ cycle when the current $i_h$ crosses zero, such that to extract the energy from the piezoelectric element twice in a vibration cycle. The voltage and current waveforms in SECE, as well as the work cycle hysteresis, are shown in Fig. \ref{fig:pvwave}(c) and (d). Denoting the phase difference between the switching instant and current zero crossing instant as $\varphi$, the original SECE is a special case of PV-SECE in which $\varphi=0$, as shown in Fig. \ref{fig:pvwave}(c). Lefeuvre et al. \cite{Lefeuvre_Badel_Richard_Guyomar_2005} introduced a different switch phase shift in SECE and realized the PV-SECE solution for broadening the energy harvesting bandwidth \cite{Lefeuvre2017, Lefeuvre2017a}. The waveforms and work cycles of the phase lead $\varphi<0$, in-phase $\varphi=0$,  and phase lag $\varphi>0$ conditions are shown in Fig. \ref{fig:pvwave}(a)--(f), respectively. By comparing the effective work cycle hysteresis, which is formed by the fundamental harmonic $v_{p,f}$ and $q$ trajectory ($q$ denotes electric charge, the integral of $i_h$), under the three $\varphi$ cases, tuning the switching phase $\varphi$ changes the obliquity and the enclosed area of the hysteric ellipses, which correspond to the electrically induced stiffness $K_e$ and damping $D_d+D_h$, respectively. The enclosed area is divided into the blue and green parts in all the hysteresis loops enclosed by the $v_p$-$q$ trajectories in Fig. \ref{fig:pvwave}. As explained in the previous impedance analysis, the extracted energy in one cycle can be divided into the dissipated (blue) and harvested (green) parts \cite{Chen2019}.

	\subsection{PV-S-SSHI and PV-P-SSHI}

	The similar phase-variable switching idea was implemented in different SSHI topologies by Hsieh et al. \cite{Hsieh2015}. The circuit topologies of (PV-)S-SSHI and (PV-)P-SSHI are shown in Fig. \ref{fig:circuits_s}(c) and (d). Compared to SECE, the SSHI solutions use an external inductor $L_i$ to realize rapid electrical resonance, the piezoelectric voltage $v_p$ across the piezoelectric capacitance $C_p$ can be flipped to a more opposite value at the current crossing zero points.  The corresponding waveforms and work cycles of PV-S-SSHI and PV-P-SSHI under different switching phase differences are shown in Fig. \ref{fig:pvwave}(g)--(r), respectively. It can be observed from the work cycle in Fig. \ref{fig:pvwave}(h) and Fig. \ref{fig:pvwave}(n) that, with a switch phase lead, i.e., $\varphi<0$, an inductive impedance is induced in the electrical part, which corresponds to the combination of an equivalent mass and equivalent damping in the mechanical side. With a switch phase lag, we get a capacitive impedance in the electrical side, which corresponds to an equivalent stiffness and equivalent damping in the mechanical side. In  PV-S-SSHI and PV-P-SSHI, there is another tunable parameter besides $\varphi$. In PV-S-SSHI, we take $\tilde{V}_r$ the rectified voltage as the other tunable parameter; while in PV-P-SSHI, we take the blocking angle $\theta$ as the other tunable parameter to better elaborate the functional relation. $\theta$ and $\tilde{V}_r$ have a one-to-one correspondence in PV-P-SSHI.

	The previous analyses of phase-variable technologies have two deficiencies. The boundary of the tunable impedance ranges was not clarified \cite{Lefeuvre2017, Hsieh2015}; the waveform description in the PV-P-SSHI was not precise because of the ignorance of the block angle $\theta$ \cite{Hsieh2015}.

	Owning to the benefit of the equivalent impedance analysis, the dynamics of different circuits under different operation conditions can be singled out and compared on the impedance plane. Fig. \ref{fig:Ze_s} shows the nine equivalent impedance points, which correspond to the phase-lead, in-phase, and phase-lag cases in PV-SCE, PV-S-SSHI, and PV-P-SSHI. The figure gives an intuitive idea about how the phase-variable technologies introduce the changes in the imaginary part of the equivalent impedance $Z_e$, which contributes to the entire system's resonance tuning to some extents. It can be roughly tell that the PV-SSHI technology has a better tunability, since the SSHI interface circuits are more capable of enhancing the effective electromechanical coupling and the backward influence to the mechanical dynamics.

	\section{Impedance Analysis}
	\label{sec:imped}

	Besides the qualitative understanding of the broadband effect of the phase-variable solutions, a quantitative study can be carried out based on the impedance analysis. The equivalent impedance provides a common tool for better comparison among the dynamic effects of different interface circuits.

	The dynamics of the power conversion circuits are unable to be quantified without combining with the piezoelectric capacitance $C_p$ \cite{Liang2012}. The impedance of the $C_p$ and interface circuit combination is denoted by $Z_e$, as given in \eqref{eq:Ze}. Harmonic analysis is carried out by firstly assuming that the current flowing through $Z_e$ is sinusoidal, i.e.,
	\begin{equation}
	i_{h}(t) = I_h \sin (\omega t),
	\label{eq:i_p}
	\end{equation}
	where $I_h$ is the magnitude of $i_h$. By defining the phase lag between the bias-flip instant and the $i_h$ zero-crossing instant as $\varphi$ and assuming each bias-flip action take much less time than a vibration cycle, the piezoelectric voltage $v_p$ can be formulated by the following piecewise equations
	\begin{equation}
	\begin{aligned}
	&{v_p}(t)=V_{oc}\times\\&
	\left\{
	\begin{aligned}
	- &\tilde{V}_1 +\cos \varphi - \cos (\omega t),&\varphi &\le \omega t<\theta  ;\\
	&\tilde{V}_1 -\cos \varphi  -\cos (\omega t),&\pi  + \varphi &\le \omega t< \pi  + \theta. \\
	\end{aligned}
	\right.
	\end{aligned}
	\label{eq:pvssshi}
	\end{equation}

	In \eqref{eq:pvssshi}, $\varphi \in [-\pi/2,\pi /2]$. Negative $\varphi$ means switching phase lead, while positive number means switching phase lag. $\tilde{V}_1=V_1/V_{oc}$ is the non-dimensionalized voltage of $V_1$, the end voltage of the positive-to-negative bias-flip actions, which is illustrated in Fig. \ref{fig:pvwave}.  $V_{oc}=I_h/(\omega C_p)$ is the nominal open-circuit voltage. The value of $\tilde{V}_1$ can be obtained by solving the following linear equations\

	\begin{equation}
	\left[\!\! {\begin{array}{*{20}{c}}
		1&1\\
		\gamma &{ - 1}\\
		\end{array}} \!\!\right]\!\!\left[\!\!{\begin{array}{*{20}{c}}
		\tilde{V}_0\\
		\tilde{V}_1\\
		\end{array}} \!\!\right] = \left[ \!\!{\begin{array}{*{20}{c}}
		{2\cos \varphi}\\
		{(\gamma-1)\tilde{V}_r}\\
		\end{array}} \!\!\right],
	\label{eq:V01_PVSSSHI}
	\end{equation}
	where $\tilde{V}_0$ is the non-dimensionalized voltage of $V_0$; $\gamma$ is the flipping factor of the voltage bias-flip actions \cite{Liang2017}.

	The waveform expression of PV-P-SSHI is more complicated than of PV-S-SSHI. The piecewise equations under the switching phase lead or lag cases are different. For the phase-lead cases, $v_p$ can be expressed as follows	\begin{equation}
	\begin{aligned}
	&{v_p}(t)=V_{oc}\times\\&
	\left\{
	\begin{aligned}
	- &\tilde{V}_1 +\cos \varphi - \cos (\omega t),&\varphi &\le \omega t<\theta  ;\\
	-&\tilde{V}_1 +\cos \varphi - \cos \theta, &\theta & \le \omega t < \pi + \varphi; \\
	&\tilde{V}_1 -\cos \varphi  -\cos (\omega t),&\pi  + \varphi &\le \omega t< \pi  + \theta; \\
	&\tilde{V}_1 -\cos \varphi  + \cos \theta ,&\pi + \theta  & \le \omega t < 2\pi  + \varphi;
	\end{aligned}
	\right.
	\end{aligned}
	\label{eq:phi<0}
	\end{equation}
	where $\varphi \in [-\pi /2,0]$ and $\theta \in [-\varphi, \pi +\varphi]$ is the blocking angle of the bridge rectifier after each bias-flip action. The value of $\tilde{V}_1$ can be obtained by solving following linear equations
	\begin{equation}
	\left[\!\! {\begin{array}{*{20}{c}}
		1&1\\
		\gamma &{ - 1}\\
		\end{array}} \!\!\right]\!\!\left[\!\!{\begin{array}{*{20}{c}}
		\tilde{V}_0\\
		\tilde{V}_1\\
		\end{array}} \!\!\right] = \left[ \!\!{\begin{array}{*{20}{c}}
		{\cos \varphi  - \cos \theta }\\
		0\\
		\end{array}} \!\!\right],
	\label{eq:V01_PVPSSHI_phi<0}
	\end{equation}
	For the phase-lag condition, $v_p$ is expressed as follows
	\begin{equation}
	\begin{aligned}
	&{v_p}(t)=V_{oc}\times\\&
	\left\{
	\begin{aligned}
	-& {\tilde{V}_1} + \cos \varphi  - \cos (\omega t),\hspace{-0.6cm}&\varphi & \le \omega t < \theta ;\\
	- &{\tilde{V}_1} + \cos \varphi  - \cos \theta  ,\hspace{-0.6cm}&\theta   & \le \omega t < \pi; \\
	-& {\tilde{V}_1} + \cos \varphi  - \cos (\omega t)- \cos \theta- 1 ,\hspace{-0.6cm}&\pi  & \le \omega t < \pi  + \varphi;\\
	&{\tilde{V}_1} - \cos \varphi  - \cos (\omega t),\hspace{-0.6cm}&\pi  + \varphi & \le \omega t < \pi  + \theta ;\\
	&{\tilde{V}_1} - \cos \varphi  + \cos \theta ,\hspace{-0.6cm}&\pi  + \theta & \le \omega t < 2\pi; \\
	&{\tilde{V}_1} - \cos \varphi - \cos (\omega t) +\cos \theta + 1 ,\hspace{-0.6cm}&2\pi  & \le \omega t < 2\pi  + \varphi;
	\end{aligned}
	\right.
	\end{aligned}
	\label{eq:phi>0}
	\end{equation}
	where $\varphi \in (0,\pi /2]$ and $\theta  \in \left[\cos^{ - 1}(2\cos\varphi  - 1),\pi \right]$. $\tilde{V}_1$ can be obtained by solving the equations as follows
	\begin{equation}
	\left[\!\! {\begin{array}{*{20}{c}}
		1&1\\
		\gamma &{ - 1}\\
		\end{array}}\!\! \right]\!\!\left[\!\! {\begin{array}{*{20}{c}}
		\tilde{V}_0\\
		\tilde{V}_1\\
		\end{array}}\!\! \right]=\left[\!\! {\begin{array}{*{20}{c}}
		{2\cos \varphi  - \cos \theta  - 1}\\
		0\\
		\end{array}}\!\! \right].
	\label{eq:V01_PVPSSHI_phi>0}
	\end{equation}

	By doing the Fourier analysis, we can derive the fundamental harmonic of $v_p$ based on the piecewise expressions in \eqref{eq:pvssshi} for PV-S-SSHI and \eqref{eq:phi<0} and \eqref{eq:phi>0} for PV-P-SSHI. The fundamental component of $v_p$ is denoted as $v_{p,f}$ \cite{Liang2012}. The $v_{p,f}$ waveforms are shown by dot lines in Fig. \ref{fig:pvwave}. Based on the harmonic analysis, the complex dynamics of the interface circuits can be equivalently formulated in terms of the equivalent impedance, which is obtained in the frequency domain as follows
	\begin{equation}
	\frac{{Z_{e}}(j\omega)}{\alpha^2} = \frac{{{V_{p,f}}(j\omega )}}{{{I_{h}}(j\omega )}}=R_h+R_d-\frac{j}{\omega C_e}.
	\label{eq:Ze}
	\end{equation}

	Comparing \eqref{eq:Ze} and \eqref{eq:Ze_mech}, the mechanical impedance $Z_e$ can be obtained by multiplying its electrical equivalent $Z_e/\alpha^2$ with $\alpha^2$, which stands for the electromechanical coupling factor. Such a dynamics equivalent is applicable for all types of interface circuits, given that their time-domain expressions of $v_p$ are available. Besides the gross effect of $Z_e$, the detailed dynamics composition, i.e., $K_e$, $D_h$, and $D_d$ in \eqref{eq:Ze_mech} can also be quantified according to the work cycle analysis \cite{Liang2012}.

	It is reasonable that interface circuits have a frequency-tuning capability to some extent, given the tunable $K_e$ component in the equivalent dynamics, whose value is a function of $\omega$, $\theta$, and $\varphi$. In particular, the switching phase lead/lag $\varphi$ causes a relatively large variation in $K_e$ for frequency tuning. Nevertheless, the frequency-tuning capabilities of different interface circuits, in terms of maximum attainable bound of $Z_e$, were not quantitatively specified in the literature.

	The PV-SSHI solutions attain the same maximum $Z_e$ magnitude when load resistance $R_l=0$ in PV-S-SSHI, which is equivalent to $\tilde{V}_r=0$ in \eqref{eq:V01_PVSSSHI}; or when $R_l=\infty$ in PV-P-SSHI, which is equivalent to $\theta=\pi+\varphi$ for the $\varphi\le 0$ case in \eqref{eq:phi<0}--\eqref{eq:V01_PVPSSHI_phi<0}, and $\theta=\pi$ for the $\varphi>0$ case in \eqref{eq:phi>0}--\eqref{eq:V01_PVPSSHI_phi>0}. In such an extreme case, the $Z_e$ bound of PV-x-SSHI (x stands for P or S) can be expressed as a function of $\omega$ and $\varphi$, i.e.,
	\begin{equation}
		{Z_{e,\max}} = \frac{\alpha^2}{{\omega {C_p}}}\left[ {\frac{2}{\pi }\frac{{1 - \gamma }}{{1 + \gamma }}\left( {1 + \cos 2\varphi  - j\sin 2\varphi } \right) - j} \right].
		\label{eq:Zepvsshi}
	\end{equation}
	With the similar approach, the $Z_e$ bound of PV-SCE can be formulated as follows
	\begin{equation}
	Z_{e,\max} = \frac{\alpha^2}{\omega C_p}\left[ \frac{2}{\pi }\left( 1+\cos 2\varphi  - j\sin 2\varphi \right) -j\right],
	\label{eq:Zepvsce}
	\end{equation}
	By non-dimensionalizing the two $Z_{e,\max}$ in \eqref{eq:Zepvsshi} and \eqref{eq:Zepvsce} with respect to $\alpha^2/(\omega C_p)$, the non-dimensionalized $Z_{e,\max}$ encloses a circular region in each solution. For PV-x-SSHI, the region centered at $\frac{2}{\pi}\frac{1-\gamma}{1+\gamma}-j$ with a radius of $\frac{2}{\pi}\frac{1-\gamma}{1+\gamma}$. For PV-SCE, the region centered at $\frac{2}{\pi}-j$ with a radius of $\frac{2}{\pi}$. The impedance ranges of PV-x-SSHI and PV-SECE are shown in Fig. \ref{fig:Ze_s}.

\begin{figure}[!t]
		\centering
		\includegraphics[width=1\columnwidth,page=9]{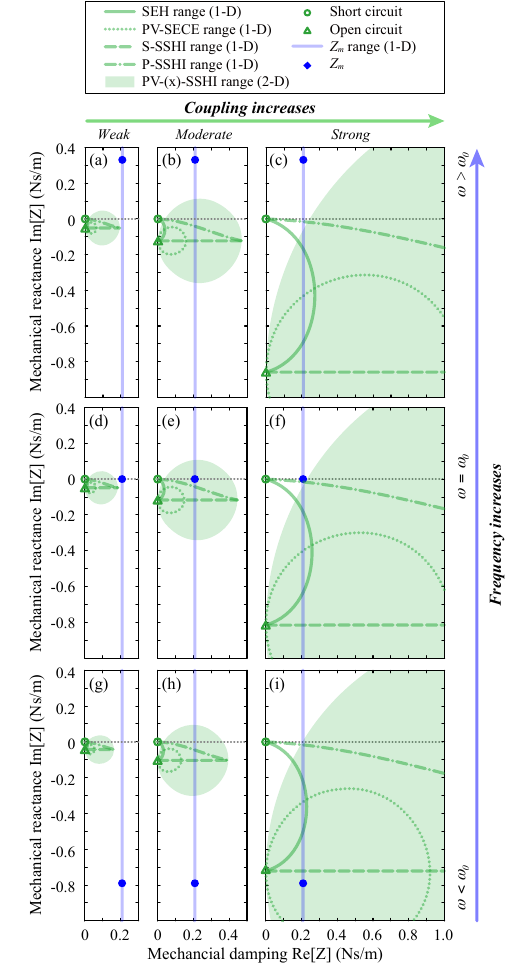}
		\caption{Impedance picture under different frequencies and coupling conditions. (a)-(c) $\omega>\omega_0$. (d)-(f) $\omega=\omega_0$. (g)-(h)  $\omega<\omega_0$. (a), (d), and (g) Weak coupling. (b), (e), and (h) Moderate coupling. (c), (f), and (i) Strong coupling.}
		\label{fig:couple}
	\end{figure}

	To implement impedance matching, the dielectric loss damping $D_p$ is firstly neglected. Its influence on impedance matching will be explained separately for simplifying the analysis of impedance matching between $Z_m$ and $Z_e$. As shown in Fig. \ref{fig:couple}, attainable mechanical impedance, and the operation point are shown as the blue curve and the blue point. The operation frequency will reduce and reach the resonance frequency at zero mechanical reactance condition from top to bottom of the mechanical impedance curve. The attainable impedance range of different circuit solutions are shown as one-dimensional green curves and two-dimensional green regions. It can be found that the attainable impedance extends to a two-dimensional circular region by the phase-variable technology, and the range of the two-dimensional impedance plane is larger than that of the one-dimensional circuit solutions.

	Fig. \ref{fig:couple} shows a case study of a certain PEH system, which is used to explain the impedance matching method under different coupling conditions and operation frequencies. Based on the operation frequency, it can be divided into three groups. The first group has an operation frequency $\omega>\omega_0$ indicated with Fig. \ref{fig:couple}(a)-(c), the second group operates at the resonance frequency $\omega_0$ shown as Fig. \ref{fig:couple}(d)-(f), and the third group has a smaller operation frequency $\omega<\omega_0$ indicated with Fig. \ref{fig:couple}(g)-(i). Based on the coupling conditions, there are also three cases corresponding to weak coupling, moderate coupling, and strong coupling conditions, respectively.

	If the PEH system operates at the resonance frequency, its mechanical impedance is an invariant value indicated as the blue point in Fig.\ref{fig:couple}(d)-(f) with the coordinate $(D_m,0)$. It can be seen that under a weak coupling condition, as shown in Fig.\ref{fig:couple}(d), the mechanical impedance point is outside the attainable impedance range of conventional interface circuits such as SEH and SECE. Compared with SEH and SCE, the in-resonance output power of P-SSHI and S-SSHI is higher because they can achieve larger equivalent damping  $D_h+D_d$, by larger load resistance $R_l$, to catch up with the mechanical damping $D_m$.  With the increase of the coupling intensity, all interface circuits can achieve larger equivalent damping as shown in Fig.\ref{fig:couple}(e) and (f). Even SEH and SCE also have the chance to match the mechanical impedance point. Especially under the strong coupling condition in Fig.\ref{fig:couple}(f), the mechanical impedance curve has two intersections with the impedance curve of SEH, which indicates two peak power frequencies that meet the impedance matching condition.  Furthermore, more advanced interface circuits such as S-SSHI and P-SSHI are no better than the conventional SEH because they all have sufficient ability to match the mechanical impedance. Therefore, the output power limit mentioned in Eq. \ref{eq:Pe_max} is not difficult to achieve under strong coupling conditions.

	For the off-resonance cases as shown in Fig. \ref{fig:couple}(a)-(c) and Fig. \ref{fig:couple}(g)-(i), although the real parts of $Z_e$ and $Z_m$ may equal to each other, the imaginary parts may not be matched for the 1-D tunable interface circuits like P-SSHI, S-SSHI or SEH since their attainable impedance ranges are 1-D curves. And for SCE, its equivalent impedance is a fixed point which is also not able to tune the imaginary part. Therefore their off-resonance and further broadband energy harvesting capabilities are limited. However, with the second tunable parameter, the synchronized switching phase $\varphi$ from the general practical PEH model,  1-D impedance curves or a fixed impedance point can become 2-D impedance regions or a 1-D circular curve, enabling a much wider range for impedance matching. As shown in Fig. \ref{fig:couple}(c), the operation frequency $\omega > \omega_0$, there is no way for 1-D tunable circuits to match the mechanical impedance point. But with a lag phase ($\varphi>0$) by PV-SSHI, which brings positive electrically induced stiffness $K_e$ into the coupling system to increase the system resonance frequency and match the mechanical impedance point. Similarly, if the operation frequency $\omega < \omega_0$ as shown in Fig. \ref{fig:couple}(i), the impedance matching can also be achieved by introducing a lead phase ($\varphi<0$) for negative electrically induced stiffness.  It's worth noting that the system coupling condition still plays an important role in impedance matching under off-resonance cases. The broadband ability can be improved by introducing strong coupling conditions or more advanced PEH interface circuits, i.e.,  the phase-variable parallel synchronized triple bais-flip (PV-P-S3BF) interface circuit\cite{Zhao2018}. With the second tunable parameter for PV-SCE and PV-SSHI, the off-resonance energy harvesting capability is increased. Therefore the phase-variable technology can broaden the bandwidth of PEH systems under off-resonance cases.

\begin{figure}[!t]
		\centering
		\includegraphics[width=1\columnwidth,page=10]{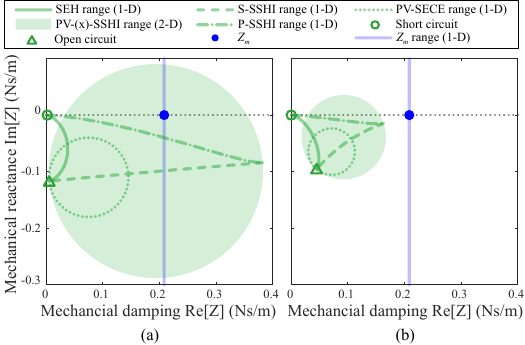}
		\caption{Influence of dielectric loss effect on impedance matching. (a) A light dielectric loss effect case, $R_p = 2M\Omega$. (b) A heavy dielectric loss effect case, $R_p = 200k\Omega$. }
		\label{fig:Rp}
	\end{figure}

	It should be noted that the dielectric loss effect also plays an important role in impedance matching. To address its influence, $R_p$ is no longer considered as infinity but realistic resistance, as shown in Fig. \ref{fig:Rp}. From Fig. \ref{fig:Rp}(a) to Fig. \ref{fig:Rp}(b), the equivalent impedance decreases with heavier dielectric loss effect, i.e., smaller $R_p$. In this case, the harvested energy from mechanical vibration is consumed by $R_p$ rather than flows to the output port. Furthermore, it can be found that the stronger the interface circuit is, the larger the dielectric influence is, and the attainable impedance ranges of SSHI topologies are decreased much more than that of SEH.  As shown in Fig. \ref{fig:Rp}(b), under a heavy dielectric loss effect, the mechanical impedance curve is far away from the equivalent impedance range, and the impedance matching can not be reached even with more advanced interface circuits since the parallel configuration between $R_p$ and the selected circuit topology as shown in Fig. \ref{fig:circuits_s}.

	\section{Experiments}
	\label{sec:experiments}

	To validate the phase variable technology, it is necessary to carry out some experiments to articulate the broadband characteristic of PV-SSHI and PV-SCE, which give some practical evidence for before mentioned impedance matching method. In this experiment, two different coupling intensities are chosen to implement different switching phases under different operation frequencies.  The improvement in energy harvesting bandwidth by using PV-SSHI and PV-SCE is checked in this experiment.

	\begin{table}[!t]
     \footnotesize
		\renewcommand{\arraystretch}{1}
		\caption{Parameters of two differently coupled PEH system.}
		\label{tab:specs1}
		\centering
		    \setlength{\tabcolsep}{1mm}{
		\begin{tabular}{cccc}
			\hline
			\bf Parameter & \bf Value  &\bf Parameter & \bf Value\\
			\hline
			\multicolumn{4}{c}{\bf Strongly coupled system} \\
			\hline
			$R$            &    24.93 k$\Omega$& $L_i$    &    47 mH\\
			$L$            &    1.61 kH & $C_r$ &    10 $\mu$F\\
			$C$            &    5.03 nF&$\omega$&    $2\pi\times55.8$ Hz\\
			$C_p$            &    22.33 nF&Acceleration    & 4.9 m/s$^2$\\
			$R_p$            &    2174.61 k$\Omega$ & Schottky diode    &    SS16\\
			$\gamma$          &    $-$0.6 & MOSFET     &    Vishay Si4590DY  \\
			$\alpha$    &  $2.35\times{10^{- 3}}$ N/V &     &  \\
			\hline
			\multicolumn{4}{c}{\bf Weakly coupled system} \\
			\hline
			$R$            &    345.47 k$\Omega$& $L_i$    &    47 mH\\
			$L$            &    31.18 kH & $C_r$ &    10 $\mu$F\\
			$C$            &    0.27 nF&$\omega$&    $2\pi\times54.8$ Hz\\
			$C_p$            &    45.7 nF&Acceleration    & 4.9 m/s$^2$\\
			$R_p$            &    1533.47 k$\Omega$ \qquad& Schottky diode    &    SS16\\
			$\gamma$ &    $-$0.6 & MOSFET     &    Vishay Si4590DY \\
			$\alpha$    &  $0.37\times{10^{- 3}}$ N/V &      &  \\
			\hline
		\end{tabular}}
	\end{table}

	\subsection{Experimental Setup}

	The phase variable interface circuit is carried out by the phase-variable control on an established prototype circuit \cite{Zhao2018}. The electronic switches are controlled by a Texas Instruments MSP430 microcontroller. Two different clamped-free piezoelectric cantilevers with strong and weak coupling conditions are achieved with different piezoelectric cover areas and beam thicknesses. They are calibrated to have similar resonance frequencies by added tip masses. The parameters of the two systems are shown in Table \ref{tab:specs1}. They are excited by a shaker under the same base acceleration. An electromagnetic sensor is mounted at the free end of the cantilever to sense the vibration velocity for the synchronized switching actions.

	\begin{figure*}[!t]
		\centering
		\includegraphics[width=2\columnwidth,page=11]{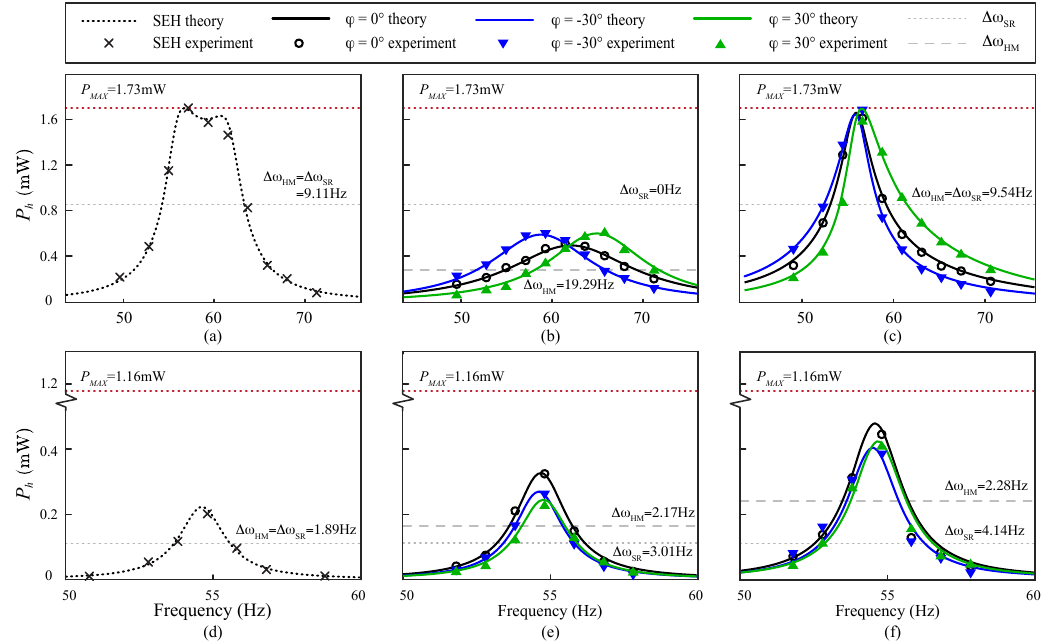}
		\caption{Harvested power $P_h$. (a)-(c) Strongly coupled system; (d)-(f) Weakly coupled system; (a) and (d): SEH; (b) and (e): PV-SCE; (c) and (f): PV-SSHI.}
		\label{fig:result}
	\end{figure*}

	\subsection{Results}

	Fig. \ref{fig:result} shows the harvested power $P_h$ under different operation frequencies and switching phase $\varphi$ for SEH, PV-SSHI, and PV-SCE. The $\varphi=0$ cases are the original P-SSHI or SCE. $\Delta \omega_{HM}$ and $\Delta \omega_{SR}$ represent the half-power bandwidth of the selected circuit and the SEH bridge rectifier referenced bandwidth, respectively. Fig. \ref{fig:result}(a)-(c) are under a strong coupling condition, the coupling points are in the attainable impedance range of SEH and PV-SSHI. Therefore they both reached the output power limit. From the detailed view of SEH shown in Fig. \ref{fig:result}(a), there are two power peak frequencies, which recall the two intersections with the mechanical impedance line in Fig. \ref{fig:couple}(f). And for PV-SCE shown in Fig. \ref{fig:result}(b),  it only has one tunable parameter $\varphi$, the electrical impedance is outside the mechanical impedance line when $\varphi=0$, which explains the reason why its power performance is inferior to SEH and PV-SSHI. However, with a lead or lag switching phase, the electrical impedance of PV-SCE can approach the mechanical coupling points. Therefore, the harvested power has been enhanced by phase variable technology. Also, its half-power bandwidth $\Delta \omega_{HM}$ is 141\%  broader than the original SCE. Compared with PV-SCE, PV-SSHI  reaches the output power limit, as shown in Fig.\ref{fig:result}(c), the $\Delta \omega_{HM}$ of PV-SSHI is 154\% broader than the original P-SSHI.

	Fig. \ref{fig:result}(d)-(f) are under a weak coupling condition. From SEH to PV-SCE and then to PV-SSHI, $P_h$ has been increased. The reason is that under the weak coupling  condition, advanced interface circuits like PV-SSHI have a larger attainable impedance range which helps to reach or get close to the mechanical impedance. Nevertheless, the bandwidth tuning ability provided by phase variable technology is confined also by the coupling condition, the broadband effect is not significant, which yields the requirement for stronger power conditioning circuits such as PV-P-S3BF. From Fig. \ref{fig:result}, under two coupling conditions, the bandwidth-broadening effect of phase variable control is successfully validated.

	\section{Conclusion}
	\label{sec:conclusion}

    A comprehensive analysis of phase-variable synchronized bias-flip technology was introduced in this paper for broadening the energy harvesting bandwidth of the piezoelectric energy harvesting (PEH) systems. Both the electrically induced damping and electrically induced stiffness/mass can be tuned in operation simultaneously. The impedance matching method is discussed to better understand and quantify the electromechanical joint dynamics and harvested power by using PV-SSHI and PV-SCE.  And the influence of the dielectric loss effect on impedance matching has been analyzed from a practical perspective. Experiments are carried out to validate the theoretical analysis. The proposed solution and analysis provide new insight into the designs of broadband as well as high energy harvesting capability PEH systems.

     \section*{Acknowledgment}

	This work was supported by the grants from National Natural Science Foundation of China (62271319 and U21B2002) and Natural Science Foundation of Shanghai (21ZR1442300).


 \bibliographystyle{elsarticle-num}
 \bibliography{reference}

\begin{thebibliography}{10}
\expandafter\ifx\csname url\endcsname\relax
  \def\url#1{\texttt{#1}}\fi
\expandafter\ifx\csname urlprefix\endcsname\relax\def\urlprefix{URL }\fi
\expandafter\ifx\csname href\endcsname\relax
  \def\href#1#2{#2} \def\path#1{#1}\fi

\bibitem{Zhao2018a}
B.~Zhao, J.~Liang, On the circuit solutions towards broadband and
  high-capability piezoelectric energy harvesting systems, in: Active and
  Passive Smart Structures and Integrated Systems {XII}, Vol. 10595,
  International Society for Optics and Photonics, 2018, p. 105950E.

\bibitem{Ng2005}
T.~H. Ng, W.~H. Liao, Sensitivity analysis and energy harvesting for a
  self-powered piezoelectric sensor, Journal of Intelligent Material Systems
  and Structures 16~(10) (2005) 785--797.

\bibitem{Cottone2012}
F.~Cottone, L.~Gammaitoni, H.~Vocca, {others}, Piezoelectric buckled beams for
  random vibration energy harvesting, Smart Materials and Structures (2012).

\bibitem{wang2019integration}
L.~Wang, T.~Tan, Z.~Yan, D.~Li, B.~Zhang, Z.~Yan, Integration of tapered beam
  and four direct-current circuits for enhanced energy harvesting from
  transverse galloping, IEEE/ASME Transactions on Mechatronics 24~(5) (2019)
  2248--2260.

\bibitem{cha2019parameter}
Y.~Cha, H.~You, Parameter study on piezoelectric length to harvesting power in
  torsional loads, IEEE/ASME Transactions on Mechatronics 24~(3) (2019)
  1220--1227.

\bibitem{Liang2017}
J.~Liang, Synchronized bias-flip interface circuits for piezoelectric energy
  harvesting enhancement: A general model and prospects, Journal of Intelligent
  Material Systems and Structures 28~(3) (2017) 339--356.

\bibitem{Szarka2012}
G.~D. Szarka, B.~H. Stark, S.~G. Burrow, Review of power conditioning for
  kinetic energy harvesting systems, IEEE Transactions on Power Electronics
  27~(2) (2012) 803--815.

\bibitem{Lefeuvre_Badel_Richard_Guyomar_2005}
E.~Lefeuvre, A.~Badel, C.~Richard, D.~Guyomar, Piezoelectric energy harvesting
  device optimization by synchronous electric charge extraction, Journal of
  intelligent material systems and structures 16~(10) (2005) 865–876.

\bibitem{Guyomar2005}
D.~Guyomar, A.~Badel, E.~Lefeuvre, C.~Richard, Toward energy harvesting using
  active materials and conversion improvement by nonlinear processing, IEEE
  Transactions on Ultrasonics, Ferroelectrics and Frequency Control 52~(4)
  (2005) 584--595.

\bibitem{Dicken_Mitcheson_Stoianov_Yeatman_2012}
J.~Dicken, P.~D. Mitcheson, I.~Stoianov, E.~M. Yeatman, Power-extraction
  circuits for piezoelectric energy harvesters in miniature and low-power
  applications, IEEE Transactions on Power Electronics 27~(11) (2012)
  4514–4529.

\bibitem{zhao2020series}
B.~Zhao, K.~Zhao, X.~Wang, J.~Liang, Z.~Chen, Series synchronized triple
  bias-flip circuit: Maximizing the usage of a single storage capacitor for
  piezoelectric energy harvesting enhancement, IEEE Transactions on Power
  Electronics 36~(6) (2020) 6787--6796.

\bibitem{Shu2007}
Y.~C. Shu, I.~C. Lien, W.~J. Wu, An improved analysis of the {SSHI} interface
  in piezoelectric energy harvesting, Smart Materials and Structures 16~(6)
  (2007) 2253--2264.

\bibitem{zhao2022ecm}
B.~Zhao, H.~R. Thomsen, J.~M. {De Ponti}, E.~Riva, B.~{Van Damme},
  A.~Bergamini, E.~Chatzi, A.~Colombi, A graded metamaterial for broadband and
  high-capability piezoelectric energy harvesting, Energy Conversion and
  Management 269 (2022) 116056.

\bibitem{zhou2019new}
G.~Zhou, Z.~Li, Z.~Zhu, B.~Hao, C.~Tang, A new piezoelectric bimorph energy
  harvester based on the vortex-induced-vibration applied in rotational
  machinery, IEEE/ASME Transactions on Mechatronics 24~(2) (2019) 700--709.

\bibitem{Tang2010}
L.~Tang, Y.~Yang, C.~K. Soh, Toward broadband vibration-based energy
  harvesting, Journal of Intelligent Material Systems and Structures 21~(18)
  (2010) 1867--1897.

\bibitem{yuan2019harmonic}
T.-C. Yuan, J.~Yang, L.-Q. Chen, A harmonic balance approach with alternating
  frequency/time domain progress for piezoelectric mechanical systems,
  Mechanical Systems and Signal Processing 120 (2019) 274--289.

\bibitem{fang2022multistability}
S.~Fang, S.~Zhou, D.~Yurchenko, T.~Yang, W.-H. Liao, Multistability phenomenon
  in signal processing, energy harvesting, composite structures, and
  metamaterials: A review, Mechanical Systems and Signal Processing 166 (2022)
  108419.

\bibitem{morel2022comparative}
A.~Morel, A.~Brenes, D.~Gibus, E.~Lefeuvre, P.~Gasnier, G.~Pillonnet, A.~Badel,
  A comparative study of electrical interfaces for tunable piezoelectric
  vibration energy harvesting, Smart Materials and Structures 31~(4) (2022)
  045016.

\bibitem{Morel2019}
A.~Morel, A.~Badel, R.~Gr{\'e}zaud, P.~Gasnier, G.~Despesse, G.~Pillonnet,
  Resistive and reactive loads' influences on highly coupled piezoelectric
  generators for wideband vibrations energy harvesting, Journal of Intelligent
  Material Systems and Structures 30~(3) (2019) 386--399.

\bibitem{gibus2022non}
D.~Gibus, P.~Gasnier, A.~Morel, N.~Garraud, A.~Badel, Non-linear losses study
  in strongly coupled piezoelectric device for broadband energy harvesting,
  Mechanical Systems and Signal Processing 165 (2022) 108370.

\bibitem{Hsieh2015}
P.~H. Hsieh, C.~H. Chen, H.~C. Chen, Improving the scavenged power of nonlinear
  piezoelectric energy harvesting interface at off-resonance by introducing
  switching delay, IEEE Transactions on Power Electronics 30~(6) (2015)
  3142--3155.

\bibitem{Li2016}
X.~Li, Y.~Feng, B.~Shao, Piezoelectirc energy harvesting system with frequency
  mismatch tolerance, {US Patent App.} 14/623,025 (Jul.~14 2016).

\bibitem{Lefeuvre2017}
E.~Lefeuvre, A.~Badel, A.~Brenes, S.~Seok, C.-S. Yoo, Power and frequency
  bandwidth improvement of piezoelectric energy harvesting devices using
  phase-shifted synchronous electric charge extraction interface circuit,
  Journal of Intelligent Material Systems and Structures 28~(20) (2017)
  2988--2995.

\bibitem{Lefeuvre2017a}
E.~Lefeuvre, A.~Badel, A.~Brenes, S.~Seok, M.~Woytasik, C.-S. Yoo, Analysis of
  piezoelectric energy harvesting system with tunable {SECE} interface, Smart
  Materials and Structures 26~(3) (2017) 035065.

\bibitem{zhao2020dual}
B.~Zhao, J.~Wang, J.~Liang, W.-H. Liao, A dual-effect solution for broadband
  piezoelectric energy harvesting, Applied Physics Letters 116~(6) (2020)
  063901.

\bibitem{brenes2020large}
A.~Brenes, A.~Morel, D.~Gibus, C.-S. Yoo, P.~Gasnier, E.~Lefeuvre, A.~Badel,
  Large-bandwidth piezoelectric energy harvesting with frequency-tuning
  synchronized electric charge extraction, Sensors and Actuators A: Physical
  302 (2020) 111759.

\bibitem{morel2018frequency}
A.~Morel, G.~Pillonnet, P.~Gasnier, E.~Lefeuvre, A.~Badel, Frequency tuning of
  piezoelectric energy harvesters thanks to a short-circuit synchronous
  electric charge extraction, Smart Materials and Structures 28~(2) (2018)
  025009.

\bibitem{Kong2010}
N.~Kong, D.~S. Ha, A.~Erturk, D.~J. Inman, Resistive impedance matching circuit
  for piezoelectric energy harvesting, Journal of Intelligent Material Systems
  and Structures 21~(13) (2010) 1293--1302.

\bibitem{Liang2012}
J.~Liang, W.-H. Liao, Impedance modeling and analysis for piezoelectric energy
  harvesting systems, IEEE/ASME Transactions on Mechatronics 17~(6) (2012)
  1145--1157.

\bibitem{Liang2014}
J.~Liang, H.~S.-H. Chung, W.-H. Liao, Dielectric loss against piezoelectric
  power harvesting, Smart Materials and Structures 23~(9) (2014) 092001.

\bibitem{Chen2019}
C.~Chen, B.~Zhao, J.~Liang, Revisit of synchronized electric charge extraction
  ({SECE}) in piezoelectric energy harvesting by using impedance modeling,
  Smart Materials and Structures 28~(10) (2019) 105053.

\bibitem{WilliamsandYates1996}
C.~B. Williams, R.~B. Yates, Analysis of a micro-electric generator for
  microsystems, Sensors and Actuators, A 52~(1) (1996) 8 -- 11.

\bibitem{Mitcheson2008}
P.~D. Mitcheson, E.~M. Yeatman, G.~K. Rao, A.~S. Holmes, T.~C. Green, Energy
  harvesting from human and machine motion for wireless electronic devices,
  Proc. IEEE 96~(9) (2008) 1457--1486.

\bibitem{Beeby2006}
S.~P. Beeby, M.~J. Tudor, N.~M. White, Energy harvesting vibration sources for
  microsystems applications, Measurement Science and Technology 17~(12) (2006)
  R175.

\bibitem{Cook-Chennault2008}
K.~A. Cook-Chennault, N.~Thambi, A.~M. Sastry, Powering {MEMS} portable
  devices---a review of non-regenerative and regenerative power supply systems
  with special emphasis on piezoelectric energy harvesting systems, Smart
  Materials and Structures 17~(4) (2008) 043001.

\bibitem{Liang2017a}
J.~Liang, C.~Ge, Y.-C. Shu, Impedance modeling of electromagnetic energy
  harvesting system using full-wave bridge rectifier, Proceedings of SPIE 10164
  (2017) 101642N--101642N--10.

\bibitem{Liang2009}
J.~R. Liang, W.~H. Liao, Piezoelectric energy harvesting and dissipation on
  structural damping, Journal of Intelligent Material Systems and Structures
  (2009).

\bibitem{Liao_Liang_2018}
Y.~Liao, J.~Liang, Maximum power, optimal load, and impedance analysis of
  piezoelectric vibration energy harvesters, Smart Materials and Structures
  27~(7) (2018) 075053.

\bibitem{Liao2019}
Y.~Liao, J.~Liang, Unified modeling, analysis and comparison of piezoelectric
  vibration energy harvesters, Mechanical Systems and Signal Processing 123
  (2019) 403 -- 425.

\bibitem{Liang2017b}
J.~Liang, C.~Ge, Y.-C. Shu, Impedance modeling of electromagnetic energy
  harvesting system using full-wave bridge rectifier, in: Active and Passive
  Smart Structures and Integrated Systems 2017, Vol. 10164, International
  Society for Optics and Photonics, 2017, p. 101642N.

\bibitem{Ottman2002}
G.~K. Ottman, H.~F. Hofmann, A.~C. Bhatt, G.~A. Lesieutre, Adaptive
  piezoelectric energy harvesting circuit for wireless remote power supply,
  IEEE Transactions on Power Electronics 17~(5) (2002) 669--676.

\bibitem{Zhao2018}
B.~Zhao, J.~Liang, Phase-variable parallel synchronized triple bias flips
  {(PV-P-S3BF)} interface circuit towards broadband piezoelectric energy
  harvesting, in: Proceedings of the 2018 IEEE International Symposium on
  Circuits and Systems (ISCAS), 2018.

\end{thebibliography}

\end{document}